# The Potential and Implications of Generative AI on HCI Education

Generative AI on HCI Education

Ahmed Kharrufa

Open Lab, School of Computing, Newcastle University, UK, ahmed.kharrufa@newcastle.ac.uk

Ian G. Johnson

Open Lab, School of Computing, Newcastle University, UK, ian.johnson2@newcastle.ac.uk

Generative AI (GAI) is impacting teaching and learning directly or indirectly across a range of subjects and disciplines. As educators, we need to understand the potential and limitations of AI in HCI education and ensure our graduating HCI students are aware of the potential and limitations of AI in HCI. In this paper, we report on the main pedagogical insights gained from the inclusion of generative AI into a 10-week undergraduate module. We designed the module to encourage student experimentation with GAI models as part of the design brief requirement and planned practical sessions and discussions. Our insights are based on replies to a survey sent out to the students after completing the module. Our key findings, for HCI educators, report on the use of AI as a persona for developing project ideas and creating resources for design, and AI as a mirror for reflecting students' understanding of key concepts and ideas and highlighting knowledge gaps. We also discuss potential pitfalls that should be considered and the need to assess students' literacies and assumptions of GAIs as pedagogical tools. Finally, we put forward the case for educators to take the opportunities GAI presents as an educational tool and be experimental, creative, and courageous in their practice. We end with a discussion of our findings in relation to the TPACK framework in HCI.

CCS CONCEPTS • Human-centered computing ~Human computer interaction (HCI)~Empirical studies in HCI

## 1 INTRODUCTION

The examination of the ramifications stemming from the rapid advancements in Generative Artificial Intelligence (GAI) technology within the realm of education has become a prominent subject of research (e.g. [1,12]). This is notably prevalent in the domain of computer science education (e.g. [2,4,5,13]), given the presence of models like Codex [17] that are trained specifically for code generation. In contrast to the extensive exploration of GAI's effects on computer science education, there exists a scarcity of published works investigating the implications of GAI on Human-Computer Interaction (HCI) education. Recent work has principally focused on investigating the broader implications of GAI on HCI [3,8,11], focused on the implications for written exams in HCI [7], evaluating the performance of current models [16], or looking at how HCI education should integrate designing-for or designing-with machine learning technologies [6].

In response, this work aims to make an early contribution as to how GAI can be utilized to support students' deeper understanding of some HCI principles while bringing to light some of its possible misuses. To keep this focus, our results and discussions do not go over the implications of GAI on assessment, as this topic requires its own dedicated investigation, or on evaluating/commenting on the quality of GAI generated output for HCI-related work (e.g. [3,8,16]). The latter point ensures that our findings are not model-dependent and are still relevant as more capable and reliable models are released.

In preparation for the delivery of our HCI and Interaction Design module, we explored the different ways to integrate GAI into the module in a meaningful way within the constraints of the module specifications already approved. Our aim was to both understand the potential and limitations of GAI in HCI education, and to ensure our graduating HCI students leave with a critical awareness of the potential and limitations of GAI on HCI as a profession. As this is a coursework-based module, we created a design brief about designing a tool that utilizes GAI to support its users in generating personas, scenarios, and requirements, and introduced GAI tools into practical classes. Moreover, we made sure to include GAI in our discussions with the students whenever relevant throughout the module. At the end of the module, we created a survey that asked students their experience of using GAI in the module and assignment as well as their thoughts on the potential for GAI in HCI practice and education based on their experiences.

In presenting our findings we contribute to the nascent discussion and debate about the role of GAI in HCI education and offer potential avenues for further exploration of *GAI as mirror* and a pedagogical tool, offering both encouragement and caution for the adoption and integration of GAI into learning and teaching practice within HCI.

## 2  METHODS

This research is based on a third year undergraduate Human-Computer Interaction and Interaction Design module within a computing science school. The module and study reported here were in English language. It is an optional 20 credit[1] module that covers the basic principles of HCI such as understanding users, design, and evaluation as well as some application areas such as social interactions, playful interactions, ubiquitous computing, tangible and natural user interfaces, The core text for the module is Helen, S., Jenny, P. and Yvonne, R., 2019. *Interaction design: beyond human-computer interaction* 5th Ed. The module is an option for computing science students on the Computing degree programme and runs alongside other compulsory computing modules on subjects such as software engineering and programming.   The two authors of this paper co-delivered this module over 10 weeks between September and December 2023.

The module is assessed through coursework only. Students are required to respond to a design brief (see Appendix B) and make two submissions: (i) a group-based presentation (3-4 students per group) reporting on field work and three early design ideas, and (ii) an individual written submission reporting on a final, individually refined design, and its rationale.

---

[1] 10.0 ECTS credits, see also Appendix A for Syllabus

The design brief gives the students an open task of designing a GAI-powered web tool to help generate personas, scenarios, and functional requirements. The tool targets non-experts in HCI and GAI, so it should provide structured scaffolding for the users. The focus for each stage is to design the interface to ensure users input the information needed for generating good and relevant output and one that can be used flexibly in the next stage. It was made clear to the students that the focus should be on the interface and user experience and not on the GAI prompts themselves (see Appendix B).

In addition to the coursework and the frequent references to GAI's potential and limitations throughout the module, we dedicated a 2-hour practical session during the third week for the students to the use GAI tools in an HCI context. After learning about personas, scenarios, and requirements in the theoretical part of the module (lectures and seminars), the practical session required students to experiment with ChatGPT, Bard, and Claude to generate personas, scenarios, and requirements. Working in groups of 3-4 students, groups were given a mini design brief and were asked to work in stages: First, generate some personas relevant to the brief, then put the generated personas in scenarios based on the brief, and finally, use these to generate a set of functional requirements (see Appendix C).

In the weeks following the end of the module, we sent all 86 students who took part in the module a request to take part in a survey. This was done after students received their final marks and feedback to ensure they provided their input without worrying about its effect on assessment. Students were told that taking part would enter them in a prize draw to win one of five £20 Amazon vouchers. Twelve students responded to the invite and filled in the survey.

The survey asked about students' previous experience of using GAI for HCI related activities, how using it in the module affected their understanding of GAI's implications on HCI as a profession and on HCI education, how they have used GAI in the practical dedicated to GAI and personas and scenarios, how they used it for the coursework, its effect on the overall engagement with the module, and its effect on a list of the high level topics covered in the module. The module designers, and authors of this paper, designed the survey questions based on domain expertise in HCI education and in AI.

Taking a grounded theory approach to analysis, each author read through the responses to the questions taking notes, writing memos and through this process generated initial categories of information supported by the data. These were discussed between authors, refined, and defined as presented in the following section, where we report on the main findings from this survey focusing on what we believe provide useful pedagogical insights to the community.

## 3 RESULTS

It is useful to note first that only one student (S1) reported having used GAI "a little" in HCI related activities before taking this module. Two other students (S3 and S5) reported using AI "a lot" for brainstorming and idea generation (but it was not clear if this was within and HCI context or in general). The rest have not used it in any HCI related context.

## 3.1 AI as a mirror: Understanding the importance of context for design

One important learning outcome for students was gaining better appreciation of the importance of details and context as a first step in any design process. Despite our emphasis on this in theory, students realized this firsthand when they were trying to generate personas and scenarios using GAI tools. Without providing enough details they ended up with similar and very shallow outputs (e.g., "*Many students produced very similar if not the same personas etc. Unless you provide very specific information the resources generated are quite generic.*", S4). S11 commented that even with specific context, it was still difficult to generate useful scenarios in contrast to generating personas, which was much easier. Some students, like S10, realized that writing good prompts was not easy and that details are necessary ("*Sometimes it was difficult to write the prompts for the AI tool*", and "*wasn't always fully accurate (probably due to the brevity of the prompts I inputted).*"). However, others like S8, found out that they got best outputs when they started adding details, incrementally evaluating the output each time: "*Basically used the same prompt (e.g., Generate persona for this app) but with more and more detail each time. At first was getting lots of 'best case' personas/scenarios, so had to manipulate it to get some that challenged the app concepts.*"

In addition to context, the engagement with GAI helped students in thinking more carefully of who they are designing for, and the consequences of the little changes they make in the input on the generated personas and scenarios (e.g., "*It certainly helped when thinking of who we were designing for, what context our tool would be used in…*", S4).

## 3.2 AI as the persona: Benefits of GAI's interactivity

While part of the brief was about designing a UI to allow its users to create good personas with the help of AI, some students took this a step further and explored the use of 'AI as the persona'. This was one of the most interesting use cases we came across as far as the use of GAI for persona generation is concerned. Some students first suggested this use case in their group presentation: They will first use the AI to create a persona, then ask it to take the role of that persona. This allows them to bring this persona to life, to ask it questions and see how it responds. They can then put that persona into a scenario and see how it responds. This use case was repeated in the survey (e.g., *"Good to have the AI generate personas and then pretend to be that persona and react to the scenario,"* S12; "*I think it really helped to solidify my understanding of design concepts (e.g., Personas) by being able to quickly generate real examples and interacting with them rather than just learning via definitions.*" (S8).

This 'role play' on the part of GAI can be an important learning tool to support deeper understanding of concepts and how they interact together - which in our case were personas, scenarios, and requirements (e.g., "I used chat GPT and copied and pasted the brief into the chat box and then asked it to come up with personas and scenarios in a positive and negative way. this worked well" and "for user personas and scenarios it helped me learn what these mean by generating them," both by S1; "We used ChatGPT and other AI models to create different personas based on a scenario. it was helpful to see these to understand how they tied in together," S5; and "I can see a great use of GAI for UX design in that the designer can gain multiple user perspectives to integrate into their design methodology." S6). References to 'positive and negative' and 'multiple user perspectives' can be used as evidence of such deeper understanding of 'users' gained through the interaction and dialogue with GAI in the design process.

### 3.3 Too much faith in AI: Students using AI for the evaluation of their designs

A use case reported by several students which merited special attention was reports of using GAI to 'evaluate' their work before submission. This implies that students believe GAI can provide reliable feedback to be used to improve, or approve of, the quality of work before submission. Example comments in this space include: "*I used it just to practice what could be improved through my design e.g. positives and negatives of current tools and how I could incorporate and improve their functionality…*" (S4); "*Having GAI has made it easier for me to evaluate each part of the design process as I go along and this informs all the other topics*" (S6); "*The only time I felt comfortable using generative AI in my coursework was to double check that I was applying concepts from the lectures correctly. For example, 'Is x feature on my home page a good example of mapping'*" (S9); "*It assisted me in evaluating the design so I could improve it and make changes based off that.*" (S10); and, "*I used generative AI to help analyse what aspects of my designs were or weren't actually matching with design brief expectations, after feeding it both images and relevant information.*" (S11).

While some students reported using GAI to gain better understanding of some concepts such as evaluation, the number of students who reported using GAI to evaluate their own work in response to several different questions raises concerns and stand out as a topic worthy of further discussion.

### 3.4 GAI and contradictions: Conflicts in students' attitudes and actions

Other points raised by the students evidenced several contradictory views and comments on both the role of GAI in HCI education and their relationship with it in their education. Despite an awareness and articulation of inherent biases in AI models, one student commented on the usefulness of GAI to overcome designer biases, stating they were: "*wary of the bias of AI before this module and this was a topic that came up quite a lot.*" But on using GAI in the module the student described the benefits as the ability of GAI to: "*generate many, varied design resources in less time. Less personal bias,*" adding in another comment that it: "*removes your own internal bias."* (S8). What is interesting here is, while the frequent discussions about bias and GAI is about the biases inherent in GAI models based on their training data, this student is referring to how GAI can help address their own personal biases.

When discussing the role of GAI in their design work, and in imagining the use of it in the future, students were protective over their creativity and the essential role of creativity in design: *"I don't think that the software is good enough to make any impactful design decisions. The AI is not coming up with radically new ideas (yet), so the designers aren't thinking too far out of the box if they are only using GAI to aid in design"* (S6). S3 was incredulous about the use of GAI in HCI professions due to a perceived inability for GAI to be creative: *"Because HCI related professions is mostly creative, I think GAI cannot fully replace that."* Students were also dismissive of GAI replacing or replicating the creativity of HCI students, stating that outputs were simplistic and lacking originality (S9) and were concerned that its use in HCI education could restrict the creativity of students. Yet, despite the emphasis of mundane tasks and the reservations and fears around creativity, GAI was often described as a tool for ideas generation including by some students who criticized GAI for its lack of creativity (e.g., S3, S6, and S7). A usage example mentioned by five students (S1, S3, S5, S6, S7) referred to brainstorming and generating ideas. When discussing how it specifically impacted their coursework design, S2 explained how it made them more creative: *"Collectively the use of GAI sparked new ideas and prompts which impacted my creativity positively"*. Others also mentioned the use of GAI to support them at the early stages of design (e.g., "*It can help brainstorm during the initial stages of planning and idea generation.*" (S3) and: "*It *helped me start the design process by providing me with a range of ideas.*" (S6). This also included bouncing ideas with AI when co-workers are not available indicating that there are cases where GAI can be used as a peer or collaborator and not just a support tool (e.g., "*it allows you to bounce ideas off someone/something more easily when no co-workers are available.*" S7).

Another contradiction came to light when one student (S10) mentioned that the way GAI was integrated in the module showed them that the use of GAI is not always considered 'plagiarism': "*It [the module] has taught me that using AI isn't always considered 'cheating' and can be beneficial when used as a tool to help assist mundane tasks within a project*". While a common theme when talking about GAI in education is that of 'cheating' and 'plagiarism', this students use of GAI helped them better understand the potential benefits of GAI in learning without worrying about all use cases being considered cheating.

### 3.5 GAI and impact on engagement

We think it is also beneficial to report on some of the feedback as to how the inclusion of GAI in the module affected the students' engagement with the module. In response to the question: "On a scale of 1-5, rate how working with AI in this module affected how engaging you found the module." 10/12 students selected 4 or 5 out of five, two selected neutral, and none selected negative impact on engagement resulting in an average of 3.92 (see Figure 1).

Six of the students who gave a rating of 4 (S1, S5, S6, S7, S8, S11) commented that this made the module more 'relevant', 'interesting', 'engaging', and/or "fun". The term "interesting" being used repeatedly in the answers and "more fun" was used twice. Other comments by students who rated 4 or over included: *"It sparked ideas and innovations that I wouldn't have thought about without the prompt from GAI"* (S2) and S4 commented that it was influential in helping them understand the potential problems with current tools and provided insights into potential improvements.

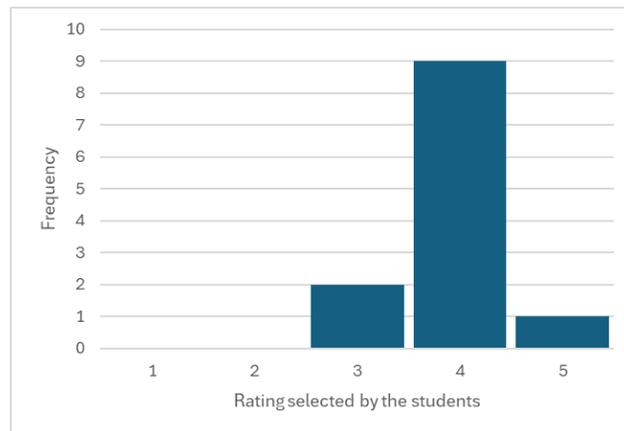

Figure 1: Student's frequency of replies to the question "On a scale of 1-5, rate how did working with AI in this module affected how engaging you found the module."

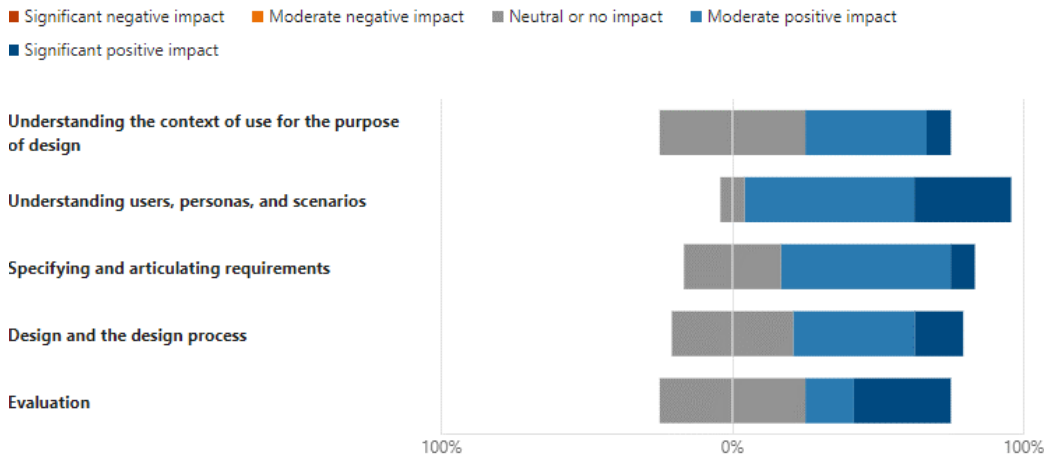

Figure 2: Students ranking of the extent to which GAI had impacted their understanding on a range of topics.

Other comments from those who gave a ranking of 4 or over included asking if the extent of the use of GAI in the module could have been pushed more (S4), as well as for asking for more exploration of the ethical considerations of using GAI in HCI (S8). One of the students who also gave 4 to this question, highlighted the importance of being aware of AI's limitations: *"As long as the students are aware of the tools limitations, and still have the ability to be creative in their work, I believe AI can make the module more engaging."* (S9).

One of the two students giving a rating of 3 (Neutral) commented that "Don't think it really changed anything about my creation process, aside from inspiring the general idea of my project." (S12). So, while GAI did not affect the engagement with the module, it did inspire the student's project.

Figure 2 shows the students answer to the question "Rank the extent to which GAI has impacted your understanding positively or negatively of the following topics." The graph shows positive impact across all topics but mainly for personas and scenarios which is understandable, as this was the focus of the coursework. But it also shows the positive impact on understanding context, requirements, the design process, and evaluation. The qualitative feedback on these points have already been reported in the subsections above.

## 4 DISCUSSION

Using a 'teaching moments' lens to reflect on the themes identified in our results allows us to discuss them through four different GAI 'opportunities': Pedagogical opportunities, education opportunities, discussion opportunities, and engagement opportunities. We then use the Technological, Pedagogical, Content, Knowledge (TPACK) framework [9] to articulate a takeaway message based on these opportunities and our experience teaching this module.

### 4.1 GAI Pedagogical opportunities

The GAI pedagogical opportunities are informed by the findings from the first two themes: AI as a mirror and AI as the persona. The integration of GAI in this module was described as 'meta' by one student in the survey alluding to the use of GAI as both 'synthetic research data' [see 8] and 'design material', also in the sense that they are using HCI concepts and learning about GAI in HCI to design a tool for part of the HCI design process that utilizes GAI. It was during the practical sessions and in our students' reflections thereof that we witnessed the most intriguing outcome. Students came into the session having had 'input' about personas and scenarios (from the lecture and associated discussions and activities) but it wasn't until they were tasked with 'prompting' GAI that the depth of their understanding and ability to articulate understanding of these concepts was tested.

The students' comments and our observations of their interaction with AI during the group-based practical sessions reveal multiple opportunities for using GAI to deliver a more constructivist, reflective, and interactive learning experience for the students. The reason we used the mirror metaphor is because the output generated by the GAI tools reflected the students' limits and level of the understanding of the topic. The concept of 'garbage in garbage out' that is common in computing science is very relevant here. When the students inputted a shallow description for the context or the requirements, they got shallow outputs that looked similar across the groups and the GAI tools used. Being able to have a dialogue and interact with the tool and seeing the impact of every change they make in the input, or adding incremental details at each interaction round played an important role in developing the students' understanding of the concepts involved.

Moreover, using GAI in groups increases the effectiveness of such interactions, as it forces students to articulate their understanding and share their critique of the output. Here, we think that carefully supported 'social interactions' [15] between students and GAI can be a productive and interesting pedagogical tool that turns the student into a teacher, in a way that encourages critical thinking, and 'deep learning' [10].

Encouraging students to use GAI to generate different types of HCI related resources, in groups whenever possible, can be a powerful tool for the educators to get the students to understand the importance of the level of understanding and details required to come up with high quality resources. Having the students interact with GAI directly and critiquing its output in their groups creates a safe, constructivist and reflective learning environment for the students. The educators' role then is to create this effective and nurturing learning environment and to ensure that students reap the maximum learning benefits made possible by this environment. Discussing the mirror metaphor with students can be a good idea to ensure that they take this critical and reflective approach to the generated output as an indicator of their understanding and how it can be developed.

### 4.2 GAI education opportunities

The number of responses from students indicating that they used GAI to evaluate the quality of their design is worrying. These responses imply a perception of GAI being more than a statistical model and attributing conceptual understanding and critical and analytical thinking characteristics to it. That we are seeing this from computing science students in their final year raises concerns about the perception and use in students less familiar with such technology.

Consequently, it is important that we educate students more explicitly about the limitations of GAI models. This was something we admittedly did not do as we overestimated the students' understanding of the technology. Asking students to experiment with such tools and using them as part of the educators' pedagogical toolkit necessitates that we make sure students know in a reasonable level of details how such tools work. Students need to understand them enough to take best advantage of them and overcome their limitations whether in the field of HCI, or even more broadly.

### 4.3 GAI discussion opportunities

Students talked about being aware of biases inherent in AI yet described using GAI tools at important parts of the design process, like generating initial ideas and evaluation. One student even commented that GAI is useful for mitigating the effects personal biases on design. In addition, students felt GAI lacked creativity and the ability to generate new ideas yet used them for these exact purposes in their own HCI education. This contradiction is summed up perfectly by S12 (see section 3.5), who played down the impact of GAI yet admitted it was the main creative spark for the project.

Such contradicting comments and conflicts between students' attitudes and perceptions of AI pose opportunities for the creation of exciting discussion activities with students about the role of GAI in HCI. Facilitating opportunities for group discussion, alongside experimentation, as part of learning and teaching activities is critical. Such discursive spaces can be configured to encourage students to hear alternative perspectives, worldviews, and positionalities as part of their professional and personal development.

### 4.4 GAI and engagement opportunities

Our survey, alongside other module feedback, showed that students found experimenting with GAIs as both design material (see Figure 1) and pedagogical tool (see Figure 2) engaging. When liberated from fears of misuse (or cheating), students discovered inspiration and an increased engagement in learning and applying key concepts and ideas from the module.

We have been enthused by this level of engagement and see opportunities to embed more pedagogical and discussion opportunities with GAI in future module development. With this paper we also hope to convince others to do the same. As such we call on educators to be bold and creative in integrating GAI in their syllabus in specific ways, moving away from a risk mitigation approach that focuses on AI as cheating toward an opportunistic approach that focuses on GAI as resource.

## 4.5 GAI and TPACK

From the data and the discussion points above, it is clear that good integration of GAI into HCI education touches on all aspects of the Technological Pedagogical Content Knowledge (TPACK) framework [9,14]. TPACK is a very popular and well-established framework particularly for educators engaged in effective teaching with technology. Through the results, analysis, and discussion, we are making the case that GAI is a technology that HCI educators need to learn enough about due to its potential role in both HCI education and HCI careers (the T in TPACK). Through the GAI pedagogical opportunities discussion, we present some possibilities into how GAI can support and enhance HCI pedagogy through the technology's mirroring, interaction, and dialogic characteristics (the P in TPACK). Finally, through the GAI education, discussion, and engagement opportunities we make the case for exploring and experimenting with integrating GAI into the HCI curriculum, whether as part of the main content, the assessment, or through discussions with and between students.

We view this paper as a starting point for a wider exploration of the potential and implications of GAI in HCI education. We also suggest that TPACK provides a good wholistic framework to guide such exploration, and that researchers already exploring the application of TPACK to HCI education such as [14] pay attention to GAI's potential on impacting all dimensions of this framework.

## ACKNOWLEDGMENTS

This research was funded by the Center for Digital Citizens (EP/T022582/1).

## REFERENCES

[1] David Baidoo-Anu and Leticia Owusu Ansah. 2023. Education in the Era of Generative Artificial Intelligence (AI): Understanding the Potential Benefits of ChatGPT in Promoting Teaching and Learning. Available at SSRN 4337484 (2023).

[2] Kharrufa Bull, Christopher Ahmed. 2023. Generative AI Assistants in Software Development Education: A vision for integrating Generative AI into Educational Practice, Not Instinctively Defending Against it. *IEEE Software* (2023).

[3] Courtni Byun, Piper Vasicek, and Kevin Seppi. 2023. Dispensing with Humans in Human-Computer Interaction Research. In *Extended Abstracts of the 2023 CHI Conference on Human Factors in Computing Systems* (*CHI EA '23*), April 19, 2023, New York, NY, USA. Association for Computing Machinery, New York, NY, USA, 1–26. . https://doi.org/10.1145/3544549.3582749

[4] Marian Daun and Jennifer Brings. 2023. How ChatGPT Will Change Software Engineering Education. In *Proceedings of the 2023 Conference on Innovation and Technology in Computer Science Education V. 1* (*ITiCSE 2023*), 2023, New York, NY, USA. Association for Computing Machinery, New York, NY, USA, 110–116. . https://doi.org/10.1145/3587102.3588815

[5] James Finnie-Ansley, Paul Denny, Brett A. Becker, Andrew Luxton-Reilly, and James Prather. 2022. The Robots Are Coming: Exploring the Implications of OpenAI Codex on Introductory Programming. In *Proceedings of the 24th Australasian Computing Education Conference* (*ACE '22*), 2022, New York, NY, USA. Association for Computing Machinery, New York, NY, USA, 10–19. . https://doi.org/10.1145/3511861.3511863

[6] Rahel Flechtner and Aeneas Stankowski. 2023. AI Is Not a Wildcard: Challenges for Integrating AI into the Design Curriculum. In *Proceedings of the 5th Annual Symposium on HCI Education* (*EduCHI '23*), 2023, New York, NY, USA. Association for Computing Machinery, New York, NY, USA, 72–77. . https://doi.org/10.1145/3587399.3587410

[7] André Pimenta Freire, Paula Christina Figueira Cardoso, and André de Lima Salgado. 2024. May We Consult ChatGPT in Our Human-Computer Interaction Written Exam? An Experience Report After a Professor Answered Yes. In *Proceedings of the XXII Brazilian Symposium on Human Factors in Computing Systems* (*IHC '23*), 2024, New York, NY, USA. Association for Computing Machinery, New York, NY, USA. . https://doi.org/10.1145/3638067.3638100

[8] Perttu Hämäläinen, Mikke Tavast, and Anton Kunnari. 2023. Evaluating Large Language Models in Generating Synthetic HCI Research Data: a Case Study. In *Proceedings of the 2023 CHI Conference on Human Factors in Computing Systems* (*CHI '23*), April 19, 2023, New York, NY, USA. Association for Computing Machinery, New York, NY, USA, 1–19. . https://doi.org/10.1145/3544548.3580688

[9] Matthew J. Koehler, Punya Mishra, and William Cain. 2013. What is Technological Pedagogical Content Knowledge (TPACK)? *Journal of Education* 193, 3 (October 2013), 13–19. https://doi.org/10.1177/002205741319300303

[10] F. Marton and R. Säljö. 1976. On Qualitative Differences in Learning: I—Outcome and Process*. *British Journal of Educational Psychology* 46, 1 (1976), 4–11. https://doi.org/10.1111/j.2044-8279.1976.tb02980.x

[11] Michael Muller, Lydia B Chilton, Anna Kantosalo, Charles Patrick Martin, and Greg Walsh. 2022. GenAICHI: Generative AI and HCI. In

*Extended Abstracts of the 2022 CHI Conference on Human Factors in Computing Systems* (*CHI EA '22*), 2022, New York, NY, USA. Association for Computing Machinery, New York, NY, USA. . https://doi.org/10.1145/3491101.3503719

[12] Lars Oestreicher. 2023. New Perspectives on Education and Examination in the Age of Artificial Intelligence. In *2023 IEEE Frontiers in Education Conference (FIE)*, 2023. 1–9. . https://doi.org/10.1109/FIE58773.2023.10342645

[13] Arun Raman and Viraj Kumar. 2022. Programming Pedagogy and Assessment in the Era of AI/ML: A Position Paper. In *Proceedings of the 15th Annual ACM India Compute Conference* (*COMPUTE '22*), 2022, New York, NY, USA. Association for Computing Machinery, New York, NY, USA, 29–34. . https://doi.org/10.1145/3561833.3561843

[14] Margault Sacre and Carine Lallemand. 2023. Applying the TPACK model to HCI Education: Relationships between Perceived Instructional Quality and Teacher Knowledge. In *Proceedings of the 5th Annual Symposium on HCI Education* (*EduCHI '23*), 2023, New York, NY, USA. Association for Computing Machinery, New York, NY, USA, 33–42. . https://doi.org/10.1145/3587399.3587402

[15] L. S. Vygotsky and Michael Cole. 1978. *Mind in Society: Development of Higher Psychological Processes*. Harvard University Press.

[16] Eric York. 2023. Evaluating ChatGPT: Generative AI in UX Design and Web Development Pedagogy. In *Proceedings of the 41st ACM International Conference on Design of Communication* (*SIGDOC '23*), 2023, New York, NY, USA. Association for Computing Machinery, New York, NY, USA, 197–201. . https://doi.org/10.1145/3615335.3623035

[17] OpenAI Codex. Retrieved February 26, 2024 from https://openai.com/blog/openai-codex

**APPENDICES**

**Appendix A: Module Syllabus**

Relevant parts of the modules outline form and syllabus.

CSC3731 Human Computer Interaction: Interaction Design
Offered for Year 2023/2024
Owning School Computing
Semesters 1
Credit Value 20
ECTS (European Credit Transfer System) 10
Aims

- To give students an understanding of Interaction Design practice and its application to a variety of application areas including ubiquitous computing, tangible interactions, and social interactions.
- To give students an understanding of relevant Interaction Design theories.
- To equip students with a skill set in practices of Interaction Design including understanding users, prototyping, and evaluation.
- To expose the students to the issues of privacy and ethics in relation to digital technologies and how they relate to specific design decisions.
- To give students experience in GUI design on multiple platforms.
- To give students experience of, and to develop skills in, working in collaborative design teams.

Outline of Syllabus

- Understanding users and user requirements
- Fieldwork for design and evaluation
- Models, theories, and frameworks of interaction design
- Design principles
- Methods for designing with and for users

- Design considerations for ubiquitous technologies, wearables, tangibles user interfaces, interactive surfaces, and natural user interfaces
- Designing for specific application areas such as user experience and social interactions.
- Prototyping
- Expert and user evaluation techniques
- Data, ethics and privacy.

Teaching Methods

- Each week, students will be provided with online learning material (short videos, slides and/or text) that will be used to introduce the learning material and for demonstrating the key concepts by example.
- Lectures sessions (2 hours/week) will provide opportunities to discuss the materials covered in the week and for live questions and answers about these topics.
- Students will be expected to read suggested material or engage with suggested online media resources in preparation for seminar sessions. Students are expected to widen their knowledge beyond the content of lecture notes and seminar material through wider self-directed background reading.
- Seminars (1 hours/week) will provide an environment to discuss published research and online material providing students opportunities to be exposed to different perspective about the covered topics and to engage in critical discussions around these topics.
- Practical classes (4 hours/week) will develop skills through hands-on experience of interaction design techniques and developing GUI design on multiple platforms. These will also promote group working skills. Students will be expected to spend significant time in completing work between practical classes, which will be used towards their coursework.

Assessment Methods

- Group showcase of early technology concepts in response to a design brief (equivalent to 1000 words, 30% of total mark)
- Individual design report work in response to a design brief (equivalent to 2500 words, 70% of total mark)

**Appendix B: Design Brief**

Relevant extracts from the design brief given to the students.

You have been given an open task of designing a generative AI-powered web tool than can help generate resources for design that will be used in the early stages of an HCI design cycle. The tool should be mainly targeting non-experts, but your design can optionally demonstrate how the tool can also adapt to support more expert users. The tool must support generating the following three outputs.

1. Personas. What information (or inputs) do you need to gather from the user to ensure generating a good and relevant persona to meet a specific context/design brief?

2. Scenarios. In addition to input and output from stage 1, what other inputs are needed from the user to generate relevant and useful scenarios that use the personas generated in stage one?
3. Initial set of functional requirements derived from the inputs and outputs of stages 1 and 2 above. What further inputs you may want from the user to generate a useful and actionable set of functional requirements.

Keep in mind that the users are non-experts so do not expect them to provide enough, or the right information needed to generate good output without enough clear guidance and structure from the interface. As such, this should not be a simple textual interface for a prompt and its output like ChatGPT but one that provides structured input (to guide non-experts in what information to provide) and provide structured output (to help non-experts make sense of the information, use it in the following stages of the process, and share it professionally). You need to make this an easy to use and enjoyable interface, so avoid relying only on 'text areas' and be more creative in how you use different UI elements to gather input from the users. You need to carefully think of the best UI element for each input and the constraints that element need to enforce. Think of what additional features are needed to make this a good, usable, useful, and possibly even enjoyable and playful product.

…

The design must be well-grounded in the principles covered in the module (e.g. principles, guidelines, heuristics, theories, experience) and ideally playful and innovative. You will start your design exploration as a group, and you will be given feedback early on in a group presentation. You are then expected to work individually and develop the idea further and explain it in your individual report...

Important note: We are interested in the interface and the experience of using it. While this involves ensuring you include interface elements that lead to gathering enough relevant information from the user about their intended design to be able to construct a prompt that can generate useful output, we are not interested in the prompts itself or the specifics of the prompt engineering that would happen behind the scenes.

**Appendix C: Week 3 Practical brief**

Relevant extracts from the practical session brief

Week 3 - Personas & Scenarios
Session objectives

- Think critically about the different GAI tools and the possible role in interaction design.
- Create set of personae and related scenarios using GAI tools.

Activity 1 (5-10 minutes): What are personae for? What are Scenarios for?
Think about what we discussed in the lectures and seminars.

- What are the different types of personae?
- What are the main reasons for creating personae? In pairs, discuss the what the most important motivations for creating personae are in interaction design.
- What is the purpose of scenarios? What makes a good scenario? In pairs, discuss what you think a good scenario looks like. What does it feature and what does it not?

Activity 2 (20-25 minutes): Experimenting with Generative AI tools and prompts to create personae.

Brief: You are part of a design team implementing a cash-less system integrated transport system in a large urban centre and surrounding sub-urban and rural areas. The new digital system will cover all forms of transport, including driver-less buses and trains, and bicycles and e-scooters for hourly hire. The focus of the new system is to make efficiencies and bring the existing system 'up to date' but must also be inclusive and accessible to all residents and visitors. For this stage of the design, you are asked to focus on the design and deployment of public screens and ticket machines that will replace existing services.

There are a wide range of GAIs to experiment with. Text-to-text Chat Bots use language models to train the AI to produce human-like responses. Some are connected to the web and that is how they have up-to-date information, while others depend solely on the information they are trained with. Of those available, some require you to sign up, some are free, and some have free versions. Research some of these available tools and make a note of the practical uses.

In pairs, or small groups (3), find and experiment with three appropriate AI tools for this activity and try three prompts for producing a persona. Take notes on each response produces in terms of its usefulness. You can try the same prompts in each tool or try different prompts. For example you could try: https://openai.com/chatgpt; https://bard.google.com/; https://claude.ai/login.

[A table to list the three prompts used with each tool tested and notes on response to each prompt]

Activity 3 (20-25 minutes): Creating personae for a specific brief using AI tools.

- After experimenting further generate 3-5 user personas for the brief - make a note of your best 2 for activity 4.
- Discuss these in your pairs why these are good/useful personae.

Activity 4 (20-25 minutes): Using GAI generated personae to generate scenarios.
Personae are useless without scenarios. They become useful when designers set up scenarios which help ask questions such as: How would this persona do this task? Could this persona do this task as it is designed?

- For one of your personae from activity 3 generate a scenario when things go well.
- For your other persona from generate a scenario when things go wrong.

Activity 5 (20-25 minutes): Using the GAI tools generate functional requirements using the personae and scenarios generated in previous activities.
Based on the outputs from activities 3 and 4, generate at least 5 Functional Requirements.